# A NOVEL REAL-TIME VIDEO AND DATA CAPTURE OF VEHICULAR ACCIDENT IN INTELLIGENT TRANSPORTATION SYSTEMS

Fekri M. Abduljalil

Department of Computer Science, Faculty of Educ., Arts, & Science, University of Sana'a, Sana'a, Yemen

## ABSTRACT

*In this paper, a novel real-time video and data capture of vehicle accident is proposed in Intelligent Transportation System (ITS). The proposed scheme solves the problem of huge storage needed for recording vehicle accident in the smart vehicle and in the remote ITS server. It works efficiently with small amount of storage size and guarantee saving accident video in secondary storage. It enables user to capture real-time video and data of running vehicle. It enables user to get vehicle accident video and data anytime anywhere. The scheme is implemented using testbed and its performance is evaluated. The results show that the proposed scheme guarantees record the vehicle accident in the ITS server. The proposed scheme has better results in comparison with full time video recording scheme.*

## 1. INTRODUCTION

Intelligent Transportation System (ITS) is the utilization of information technology to enhance the transportation system. Elements within the transportation system such as vehicles, roads, traffic lights, and message signs, become smart devices. ITS enable various users to be more coordinated, better informed, and smarter use of transport networks [1]. Real-time data capture in the Intelligent Transportation System is to enable the active acquisition and systematic provision of integrated, multisource data to enhance current operational practices and transform surface transportation systems management. It addresses the capture, quality-checking, and integration of real-time data from wirelessly connected vehicles, travelers, and infrastructure. Furthermore, it uses these data to develop and deploy enhanced or transformative mobility distributed applications [14]. As a result for the advances in wireless communications technologies and the availability of Internet access anywhere anytime, Automobile manufacturers continue to incorporate more and more technological features such as wireless access into their automobiles, and new applications can be developed that leads for next generation of Smart vehicle [2]. One of the most capabilities of these smart vehicles is efficient video recording of accident.

The Intelligent Transportation system must offer efficient vehicular accident management distributed system [12]. One of the most capabilities of this distributed system is efficient video recording of accident. Every year many accidents happen causing injuries and fatalities. For example, the road injury statistics in Yemen from 2001 to 2010 indicate that around 129,946 vehicles accident, 166,744 people incurred serious injuries, and 25,441 people die [3]. The traffic statistics are worse in the modern countries and it is in increase. Another example, the road injury statistics in Canada in 1996 indicate that around 230,000 people incurred serious injuries, and road fatalities were around 3,000[4]. The delay associated with these accidents increase the number of death. These statistics raise the query to achieve better accident management such as in countries with no infrastructure such as far areas. The smart vehicle should be able to record all information associated with vehicle accident. This information may include vehicle speed, the vehicle performance, vehicle operating status such as turning light condition, braking condition etc. The recorded information associated with accident could provide objective data for crash reconstruction, roadway design, driver training, psychology analysis and insurance rates settings for participating vehicles. This information becomes important evidences for finding the true situation of a traffic accident.

Two design goals of accident management are: 1) Fast accident detection to decrease the delay associated with saving injuries and give them medical help. 2) Record the information associated with accident: The information recorded before and during the accident can be used by police investigator to understand the accident cause. It also can be used as evidence on law enforcement when a traffic accident occurred [5, 6].

All previous research works indicate that video recording the information associated with accident requires huge storage. The problem is how to guarantee record the accident video and data with any available storage space. Video recording from vehicle switch on to time of accident needs huge storage space and data may loss during the accident.

In this paper, we proposed a novel scheme for a video recording for the last period before and during accident. The proposed scheme is the solution to the problem of limited storage space such that it guarantees recording video for accident with small amount of storage space.

The remainder of this paper is organized as follows. In Section 2, related works are surveyed. The proposed schemes are introduced in Section 3. Section 4 presents the system implementation details of the proposed schemes. Section 5 presents the results and discussion. Section 6 is conclusions.

## 2. RELATED WORKS

In [7], a prototype for a short recording time using a video camera with solid state memory is presented. The video camera records continuously until it senses an accident, then it stops recording and saves a video to solid state memory.

The IEEE 802.11 standard body is currently working on a new IEEE standard for wireless access in a vehicular environment (WAVE) which it is called IEEE 802.11p. The main purpose of this standard is to enable public safety applications that can save lives and improve traffic flow [8,9].
A new approach for incident identification and response system is proposed in [10]. In this research work, a statistical approach for modelling congestion associated with freeway is developed. It showed that quantifiable relationships exist between the congestion metrics and incident severity data.

In [11], An Implementation of a Vehicular Digital Video Recorder System (VDVRS) is proposed. The storage needed is 1.87GB per minute for uncompressed data full color image. The VDVR system starts recording by user and stop recording by user. It records vehicle speed and location along with video recording.
In [12], a novel framework for vehicle accident management suitable for next generation wireless networks is presented. It presents two methods for accident detection for any type of accidents. It also presents data video recording schemes to be used for accident investigation.

Different from the aforementioned works, we propose an efficient video and data recording schemes that enables ITS center to guarantee vehicle accident video and data recording using any storage size. To the best of our knowledge, this is the first approach to address real-time video and data capture of vehicle accident in intelligent transportation system. Furthermore, the proposed scheme is the first to solve the problem of huge storage size needed by video recording systems.

## 3. THE PROPOSED SCHEME

In this section the proposed scheme is presented in three subsections. The first subsection presents system architecture of the proposed scheme. The second subsection describes In-Vehicle Video and data Recording Scheme (IVVDR). The last subsection describes details of the real-time video and data capture scheme (RTVDC).

### 3.1. System Model

The scheme is proposed for system architecture presented in [12]. The system model consists of three parts which are the vehicle system, the server, and the user. The system model consists of computer system connected to different intelligent sensors and control units, which are implemented by micro-controllers as shown in figure 1. The computer system (on-board PC) connected to (1) digital camera, (2) wireless interfaces that support either one or all wireless networks, and (3) sensors connected to computer system, and (4) GPS system. The computer system records data collected from different sensors in parallel with video data recording. The computer system must be kept in strong container. The vehicle Computer system must be connected to sensors such as Air Bag sensor, and Vehicle angle sensor.

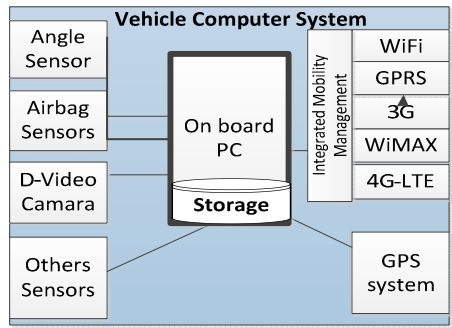

Figure 1: The proposed in-vehicle computer system.

The software part of the proposed architecture is shown in figure 2. It includes eight sub-modules, namely, Check Sensors Information, Get Vehicle Location Information, Send Vehicle Location Information, Send SMS Message, Show Vehicle Location on Map, Show Last Video, Create New Video, and Save Video. The functional requirements for the implemented system are depicted in Figure 2.

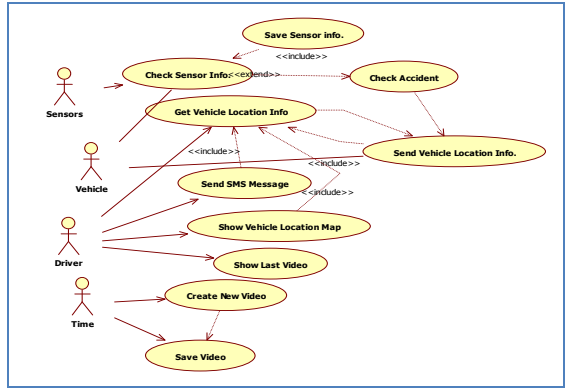

Figure 2: Use case diagram for Vehicle Computer subsystem

There are two methods for accident detection. The first one is when the vehicle turnover. We call it vehicle turnover detection (VTD). In this case, the computer system continuously senses the vehicle angle. The second detection method is called vehicle crash detection (VCD). In this method, the computer system continuously senses the air bag status [12].

Figure 3 shows system model and architecture for real-time video and data capture. The vehicle captures real time video and data and saves it in its secondary storage and sends it to server. The server receives captured video and data and saves them in its storage. The user can access real-time video and data through request server. The system model can be used by traffic police or ITS Centre or any interested company such as car rent company.

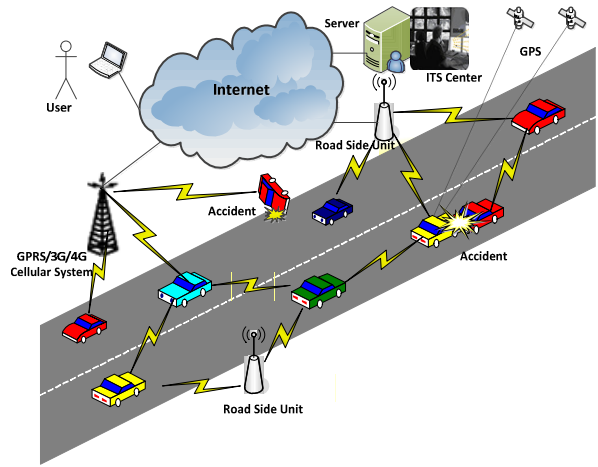

Figure 3: System Model

## 3.2. In-Vehicle Video and Data Recording scheme (IVVDR)

The objective of the proposed schemes is to record the video of accident. The most important period of video is the last minutes before and during the accident. We proposed two schemes for recording accident: The first scheme is to start recoding video from vehicle start on until vehicle stop or accident occurs. The first scheme is simple such that it records everything about that accident. This scheme needs a huge storage so that it is expensive. The second scheme is to use two video files. It starts recording in one file for T minutes, then starts record in the second file for T minutes. After the T minutes of recording in the second file is finished, it removes the contents of the first file and starts recording to it. Then the above steps are repeated until accident occurs. A timer is used to alternate recording video between the two video files. The following steps show the proposed scheme. It consists of two parts.

## **Part1 Vehicle Start On.**

```
Part 1 vehicle_start_on
Begin
...
Set Timer1.Enabled = True
'Set interval for timer "tick"
Set Timer1.Interval = 300000;
Set file_alternat=1;
Set Accident_flag=false;
Set capture1.Filename  = "e:\test1.avi";
Set capture2.Filename = "e:\test2.avi";
...
End
```

## **Part 2 variant 1: Video Recording until accident occur**

```
Part 2 On_Accident
Begin
...
Set Accident_flag=true;
'Stop Video Recording;
Set Timer1.enabled=false;
If (file_alternat==1)Then
   capture1.Stop();
else
   capture2.Stop();
End If
,other functions in the proposed system
Get GPS coordainate.
Stop Sensors data recording;
Send Message to Traffic Server IP;
Send Help SMS;
...
End
```

```
private void timer1_Tick()
Begin
  if (file_alternat == 2)
     Capture2.Stop();
     Capture1.Start();
     file_alternat = 1;
  else
      Capture1.Stop();
      Capture2.Start();
      file_alternat = 2;
  End if
End
```

**Part 2 variant 2: Video Recording Before and After Accident**

```
Part 2  On_Accident                  private void timer1_Tick()
Begin                                Begin
...                                   If (accident_flag)
Set Accident_flag=true;                  Timer1.enabled=false;
,other functions in the proposed system  if (file_alternat == 2)
Get GPS coordainate.                        Capture2.Stop();
Stop Sensors data recording;             Else
Send Message to Traffic Server IP;          Capture1.Stop();
Send Help SMS;                           End if
...                                  else  // not accident
End                                      if (file_alternat == 2)
                                            Capture2.Stop();
                                            Capture1.Start();
                                            file_alternat = 1;
                                         else
                                            Capture1.Stop();
                                            capture2.Start();
                                            file_alternat = 2;
                                      End if
                                     End if
                                     End
```

The first part (part 1) is executed when vehicle starts running. It enables the timer to initiate video recording and it sets timer interval to 300,000 milliseconds. This time interval is equal to 5 minutes. It can be decreased or increased according to two factors which are the storage available and the period of video recording. The timer1_tick() is invoked every period according to value specified in Timer1.Interval. In part 1, two instances of class capture are used which are capture1 and capture2. The class capture is implemented using DirectShow API [13]. It has different member methods and member variables. The member method Start() of class capture is used to initiate starting video recording to AVI file indicated in the member variable Filename. The member method Stop() of class capture is used to stop video recording to the AVI file which is specified in the member variable Filename. Part 1 uses two AVI files for video recording, the two files are test1.avi and test2.avi.

The part 2 is executed on receiving signal from vehicle sensors. There are two variant of part 2. The first variant is the Video Recording until accident occurs. When accident occurs, it disables the timer and stop video recording to the AVI file. It saves last data received from sensors and stop recording any data from any sensors. The second variant is for video recording before and after accident for limited period. It just set the global variable Accident_flag to true. It saves last data received from sensors and stop recording any data from any sensors except video recording. Because The timer1_tick() is invoked every period according to value specified in Timer1.Interval, the second variant, the scheme wait until time1_tick() is invoked, then the timer is disabled and video recording is stopped.

## 3.3. Real-time video and data capture scheme (RTVDC)

The current research of Research and Innovative Technology Administration (RITA) [14] focuses on real-time data capture and management. Real-time data capture and management is the access to high-quality real-time multi-modal transportation data that is captured from connected vehicles and infrastructure. The full time video recording scheme is not suitable for distributed system because of video data needs huge storage. The In-Vehicle Video Recording scheme is proposed to be used to record video and data inside vehicle.

In this section, a real-time video data is captured from vehicles and transmitted to the ITS remote server. The ITS remote server records accident video to its secondary storage in the same way as in the proposed in-vehicle video recording scheme.

In this section, we presented an enhancement to in-vehicle video recording scheme in order to support real time video data capture to ITS remote server. It is called RTVDC. The ITS remote server makes request for video capture to vehicle. The vehicle computer system sends video stream to remote server through the integrated wireless networks which proposed in [12]. The remote server receives video stream and records it to secondary storage according to in-vehicle video recording scheme. In this scheme, real-time data from a vehicle sensors such as speed, location GPS information and video are simultaneously sent to the remote server. This scheme is proposed to enable traffic police server to capture accident video data. We argue that the proposed in-vehicle video recording scheme is the best scheme for real-time video data capture. The remote server can be Traffic Police server, or ITS traffic management Centre, or any private/Government center that is used for vehicle traffic management.

The proposed vehicle algorithm is summarized in the following steps:

Vehicle Algorithm

Step 1: the driver starts vehicle.
Step 2: the computer system starts up.
Step 3: the vehicle automatically start video recording to its storage (run IVVRS).
Step 4: vehicle makes connection (control) with server
Step 5: vehicle informs server of its running.
Step 6: vehicle listens for a request from server.
Step 7: when server request received then
      Create two connections to the server (video stream and data stream).
      Repeat using timers
        Vehicle sends captured video stream through video stream connection.
        Vehicle sends captured vehicle data through data stream connection.
      Until vehicle stopped or accident occur
 Step 8: Send terminate report

The vehicle system establishes three UDP connections with different port numbers and same IP address to ITS Server. The three connections are control, video, and data stream. In step 5, the vehicle sends its login information. The server should be running all the time. If the server accepts the request of the vehicle, it adds vehicle to online vehicle list and sends request for video and data stream. Vehicle creates two connection video and data. Vehicle continues sends video and data captured to server based on two different timers. If vehicle stopped or accident occurs, Vehicle sends termination report to server. The proposed ITS server is summarized in the following algorithm:

ITS Centre Server Algorithm
Start-up server
Manual vehicle/user registration
Server creates five UDP connections with different ports.
While (true) {
      Server listens for vehicle ID connection request
      If server receives vehicle connection request Then

```
        If vehicle is registered vehicle Then
            It adds vehicle to list of active (running) vehicle.
            Set status to running
            Server sends request to vehicle
            When server receives real-time video data stream then
            Repeat
                Run EDVRS on this stream.
                If vehicle user is enabled then
                    Forward video and data to user
                End if
            Until vehicle terminate report received
        End if
    Else if server receive user enable request then
        If user is registered and its vehicle is running Then
            It adds user to list of active (running) user.
            It set user to be enabled.
        Else If user is registered and its vehicle is NOT running Then
            Server sends video and data stream of last received video and data.
        End if
    Else if server receive user disable request Then
        If user is registered Then
            It remove user from list of active (running) user.
            It set user to be disabled.
        End if
        End if
    End if
End while
```

There are five types of UDP connections. One UDP connection is used for control. Two UDP connections are used to receive video and data from vehicle. Another two UDP connections are used to send video and data to user. The system can intelligently detect any vehicle crash to other using location.

## 3.4. Security

The use of standard encryption techniques to encrypt entire real-time video streams adds too much computational overhead because huge data stream is encrypted, transmitted and decrypted. MPEG encoding/decoding is a computationally intensive process; therefore further research work is required to securing video data with minimal complexity for encryption and decryption.

# 4. SCHEME IMPLEMENTATION

Main objective of the proposed schemes is to record video and data during accident in the vehicle secondary storage and record video and data captured in the server. It enables user to get real time video and data of the vehicle and vehicle accident if any. This is achieved by means of an outdoor wireless LAN connected to Ethernet LAN, one personal computer as ITS Centre server, one laptop as vehicle client, one laptop as user client, and one computer as GPS server.

Figure 4 shows a logical view of how all the components interact with each other while providing framework for real time video and data capture. Each node has one interface. Each node gets configured manually as shown in figure 4. The laptop computer of vehicle is configured with the IP address of server and port numbers. The server is configured with IP address and three port numbers as shown in figure 4.

Main objective of the first proposed scheme is to record video and data during accident. This is achieved by means of 1 laptop computer connected to digital video camera inside vehicle.
The system is implemented using Microsoft Visual studio .Net 2008 using C# as an enhancement to the system implemented in [12]. Microsoft SQL server is used for creating database at server. The proposed schemes are implemented using Microsoft DirectShow application programming interface (API) [13]. DirectShow is used for capture and playback high-quality video and audio. A digital web camera is used in front of the vehicle to provide the scene image. The video frame rate is 30 frames per second. The resolution of each frame image is 640 x 480 pixels. Audio device is 2HD. The Audio sampling rate is 44.1 kHz. The sample size is 16 bit.

The connection between the vehicle computer system and the remote server is established using UDP protocol. Wireless LAN 802.11 with 2 access points (outdoor antenna and indoor antenna) is used to connect vehicles with remote server. The remote server can receive many video streams over UDP connections at the same time and then save it to secondary storage. The video stream data is sent as images. These images should be received in order as they are sent with minimum delay. The data captured is sent as text using another UDP connection with specific format. Other function of the proposed system has been simulated. The system generate random value of the vehicle angle, it continuously checks this values. If the generated value is greater than Ø degree, it invokes on accident algorithm. The Ø is the critical angle degree for vehicle turnover. The critical angle degree is different for the different type of vehicles. The velocity of vehicle is generated using the normal (Gaussian) distribution [15].

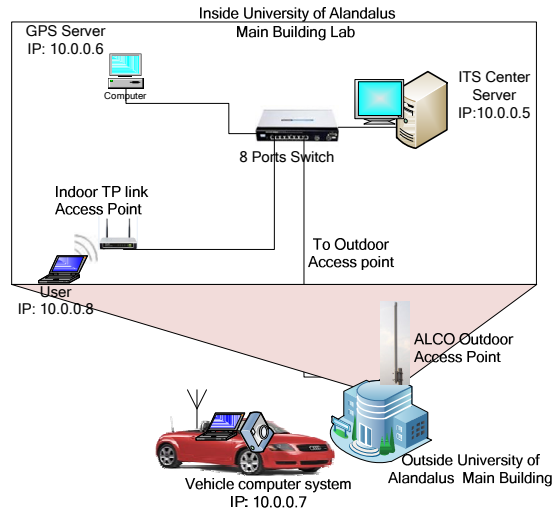

Figure 4: System testbed and components

There are many scenarios of running. The first one is to test recording the video for different time using full time video record. The second scenarios is to test the in-vehicle video and data accident recording with different recording time. The third one is to test real-time video and data capture of accident.

# 5. THE EXPERIMENTAL RESULTS AND DISCUSSIONS

In this section, we show the experimental results of running the proposed schemes. We run the proposed in-vehicle video and data recording scheme in two different scenarios. The first scenario is the full video recording. In this scenario we used only one file. The timer is not used. The system stop recording when vehicle is either make accident or storage limit is reached or vehicle stopped running.

Figure 5 shows the storage space needed by the scheme VDVRS implemented in [11] and the storage space needed by our proposed scheme. It can be observed that recording video data needs huge data storage in both schemes. The VDVRS scheme needs 1.87GB of storage for 1 minutes of uncompressed video recording with frame size of 30 fps and the resolution of each frame image is 720 x 480 pixels. Our proposed scheme needs 22.9 MB of storage for 1 minutes of uncompressed video recording with frame size of 30 fps and the resolution of each frame image is 640 x 480 pixels.

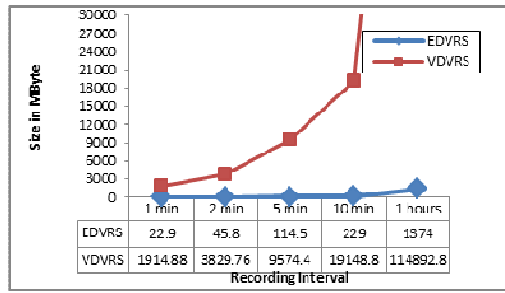

Figure 5: The storage space needed by VDVRS and IVVRS

The second scenario is the video recording before and during accident. In this scenario, the simulation (driving) time is one hours. The recording time to secondary storage of one file is 5 minutes. The max and min storage needed is reported in figure 6. The minimum storage is the size of one file, which it is guaranteed to have video and data of the accident. We run scheme in two different variants. In the first run, video recording is stopped when accident detected. The maximum storage is the size of the two video files. The total video storage is a value between the max and min. it depends on the time of accident. In the second run, video recording is stopped when recording time to current file is elapsed.

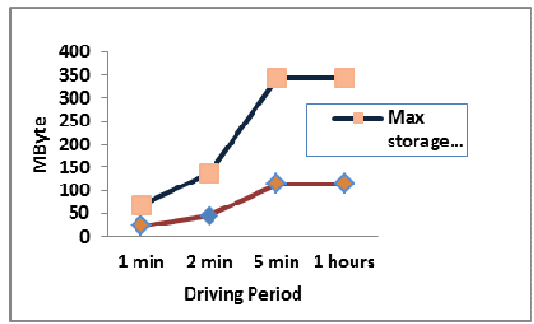

Figure 6: The maximum and minimum storage space for different driving time

Figure 6 show that the maximum and minimum storage needed by the proposed in-vehicle video recording scheme for different duration of times. The proposed scheme does not exceed the specified limit of storage and in the same time it guarantees record the accident. The scheme proposed in [11] needs unlimited storage to continue working and if the storage limit is reached it needs user to delete the video file recorded and request new recording. If the accident occurs, it continues recording until the storage becomes full. Figure 6 shows that for 2 minutes of video recording, vehicle needs 46MB of data storage. This proves that full time data recording is not suitable method for accident recording. If we set our scheme to record the last 5 minutes before the accident, vehicle needs storage size less than 114.5MB.

For further works, details of the proposed protocol that enable user to access real time video and data of vehicle will be published in future. The performance will be tested using different numbers of real time video accesses. The proposed schemes can be improved by using two video cameras: one for video recording inside the vehicle and another one for recording outside (in front of) the vehicle. Our proposed scheme is more scalable which support accident video recording with any storage space and has the ability to limit the storage space needed to record video for accident.

## 3. CONCLUSIONS

In this paper, a novel enhancement to in-vehicle digital video recording scheme of smart vehicle is proposed. The proposed scheme supports real-time video and data capture of running vehicles in the intelligent transport system. The contributions of this paper are that it solves the problem of huge storage required by previous video data recording schemes. It efficiently guarantees records video before and during the accident in the remote ITS server. It needs a limited storage space and this storage space can be controlled. Part of the proposed scheme is implemented using test bed. Another part of proposed scheme is simulated. It is found that the proposed scheme is scalable and efficient for recording video and data of accident.

## ACKNOWLEDGEMENTS

A substantial part of this research was financed by Alandalus University, Republic of Yemen. I am grateful to University of Alandalus for having believed that this might be a real and reasonable investment.

**Authors**

FEKRI M. A. ABDULJALIL  (fekri@andalusuniv.net) received a B.Sc.degree in computer science from Baghdad University, Iraq, in 1997 and a Master of Computer Science degree from University of Pune, India, in 2002 and a Ph.D. degree in 2008, from Pune University, Pune, India. Currently, he is an assistant professor at the dept. of mathematics and the head of the dept. of computer science, Sana'a University. His research area of interests includes computer networks, mobile computing, IP mobility protocols, Mobile Ad hoc Routing protocols, Analysis and design of protocols, Vehicular Networks, and Intelligent Transportation System.

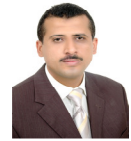